\documentstyle[12pt]{article}
\begin{document}
\topmargin 0pt
\oddsidemargin 7mm
\headheight 0pt
\topskip 0mm

\addtolength{\baselineskip}{0.40\baselineskip}

\hfill SNUTP 94-14

\hfill February 1994

\begin{center}

\vspace{36pt}
{\large \bf Schr\"{o}dinger Fields on the Plane with non-Abelian
Chern-Simons Interactions}

\end{center}

\vspace{36pt}

\begin{center}

Won Tae Kim$^*$ and Choonkyu Lee

\vspace{20pt}

{\it Center for Theoretical Physics and Department of Physics, \\
Seoul National University, Seoul 151-742, Korea}

\end{center}

\vspace{10pt}

\vfill

\begin{center}
{\bf ABSTRACT}
\end{center}
\hspace{1cm}
Physical content of the nonrelativistic quantum field theory with non-Abelian
Chern-Simons interactions is clarified with the help of the equivalent first-
quantized description which we derive in any physical gauge.
\vspace{12pt}

\begin{center}
(Submitted to: Physical Review D)
\end{center}

\noindent

\vspace{24pt}

\vfill

-------------------------------------------------------------------- \\
$^*$E-mail address: wtkim@phya.snu.ac.kr \\

\newpage
\section{INTRODUCTION}
\hspace{1cm}
Quantal structure of the nonrelativistic field theory on the
plane with Abelian Chern-Simons interactions has been discussed
recently in Ref. \cite{jp} : it reduces to the theory of particles obeying
exotic Abelian statistics, usually called anyons \cite{any}. In
two spatial
dimensions, it is known that particles obeying non-Abelian braid
statistics are also possible \cite{fro}. The latter case is realized if
particles are allowed to couple to statistical gauge fields whose dynamics
is governed by the non-Abelian Chern-Simons Lagrange density \cite{djt}
\begin{equation}
\label{eq:cs}
{\cal L}_{CS} =-\kappa \epsilon^{\mu \nu \lambda} tr \left(
A_\mu \partial_\nu A_\lambda +\frac{2}{3} A_\mu A_\nu A_\lambda \right),
\end{equation}
where $\kappa$ is a dimensionless constant and $A_\mu =A^a_\mu T^a $
is the matrix-valed gauge connection. (The $T^a$'s are antihermitian group
generators satisfying $[T^a,T^b]=f^{abc} T^c$ and $tr(T^a T^b)=-\frac{1}{2}
\delta^{ab}$). Gauge invariance of the theory requires quantization of the
parameter $\kappa$, $\kappa=\frac{n}{4\pi} (n \in Z)$ \cite{ann}.

Quantum mechanical or first-quantized description of these
non-Abelian Chern-Simons particles has been given in the so-called
holomorphic {\it gauge} \cite{vel,ltj}.
In more usual (e.g. Coulomb) gauges, the nonlinear
nature of the Gauss law has been a major obstacle to the explicit
treatment. As an alternative formulation, one can also contemplate
the field-theoretic or second-quantized description of these particles,
i.e. the non-Abelian extension of the treatment given in
Ref. \cite{jp}. In this paper, we present such a treatment
and show explicitly that this system has an equivalent first-quantized
description with all the desired properties.
Note that, when one posits
the second-quantized Hamiltonian, the holomorphic gauge choice
becomes problematic since the corresponding Hamiltonian possesses
no obvious Hermiticity property. But for our purpose at least,
any gauge choice preserving the Hermiticity of the vector potentials
$A^a_i$ will do.
In the next section, we will specify our field theory system
and then derive the many-body Schr\"{o}dinger equation implied
by it. The last section is devoted to the discussion of our
finding.

\section{NONRELATIVISTIC FIELD THEORY AND ITS FIRST-QUANTIZED DESCRIPTION}
\hspace{1cm}
The second-quantized description utilizes the (complex) Schr\"{o}dinger
fields $\psi_\alpha ({\bf r},t)$ which interact minimally with non-Abelian
Chern-Simons gauge fields $A^a_\mu$. Here, ${\bf r}=(x, y)$ and
$\alpha$ denotes the internal index associated with a specific
representation of the gauge group appropriate to given particles.
We work with Heisenberg-picture field operators.
Then, based on the standard Lagrangian density which includes ${\cal L}_{cs}$
given above, we readily obtain the Hamiltonian of the system
(we set $\hbar =1$ and $i=1, 2$),
\begin{eqnarray}
\label{eq:ham}
H & = & \frac{1}{2m} \int d^2 {\bf r} \psi^\dagger ({\bf r},t) (\partial_i
{\hspace{-0.1in}}^{^{^{\bf{\leftarrow}}}}
  -A^a_i ({\bf r},t) T^a ) (\partial_i +A^b_i ({\bf r},t) T^b) \psi({\bf r},t)
\nonumber \\
 &\equiv&\frac{1}{2m} \int d^2 {\bf r} (D_i \psi)^\dagger ({\bf r},t) (D_i
\psi)({\bf r},t)
\end{eqnarray}
together with the Gauss laws which relate the operators
$A^a_i ({\bf r},t)$ to non-Abelian
matter densities $\rho^a ({\bf r},t)=i \psi^\dagger ({\bf r},t) T^a \psi
({\bf r},t)$:
\begin{equation}
\label{eq:gau}
\frac{1}{2} \epsilon^{ij} F_{ij}^a \equiv  \epsilon^{ij}
\partial_i A^a_j + \frac{1}{2} \epsilon^{ij} f^{abc} A^b_i A^c_j
= -\frac{1}{\kappa} \rho^a
\end{equation}
If one wishes, one may also include in the Hamiltonian additional
gauge-invariant
interactions involving matter fields only. The Schr\"{o}dinger field operators
$\psi_\alpha$ and $\psi^\dagger_\alpha$ satisfy the usual
equal-time commutation relations
\begin{eqnarray}
\label{eq:com}
\left[ \psi_\alpha ({\bf r},t), \psi_\beta ({\bf r}',t) \right] &=&
\left[\psi_\alpha^\dagger
({\bf r},t), \psi^\dagger_\beta ({\bf r}',t) \right] =0, \nonumber \\
\left[ \psi_\alpha ({\bf r},t), \psi_\beta^\dagger ({\bf r}',t) \right] &=&
\delta_{\alpha  \beta} \delta^2 ({\bf r} -{\bf r}' ).
\end{eqnarray}
(We take a bosonic algebra for definiteness; similar analysis can be given
with a fermionic algebra.)  A simple consequence of these commutation relations
is: $\left[ \rho^a ({\bf r},t), \right. $ $\left. \rho^b ({\bf r}',t) \right]
=i f^{abc} \delta^2 ({\bf r} -{\bf r}') \rho^c ({\bf r},t)$.
We have now specified our system completely.

Does this nonrelativistic quantum field theory
describe the same physical system as that discussed in Refs. \cite{vel,ltj} ?
This can be answered by deducing the equivalent first-quantized
description. What makes the problem nontrivial (say, compared
to the Abelian case
\cite{jp,lee}) is that, in the present case, the operators
$A^a_i ({\bf r},t)$ are specified only implicitly through Eq. (\ref{eq:gau}).
We also need a properly ordered second-quantized Hamiltonian.
It turns out that the `correct' form of the Hamiltonian operator is given
by our expression (\ref{eq:ham}), with various operators ordered precisely
in the form written there. This gives the correct (Hermitian) Hamiltonian
in the sense that the most natural first-quantized
description then follows.

The source of difficulty with Eq. (3) is
of course the nonlinear term in the non-Abelian field strength $F_{ij}^a$.
(This nonlinear term would drop out in the axial gauge ($A^a_1 \equiv 0$),
but it is
not so in more {\it safe} gauges where the fields $A^a_i$ may be
assumed to vanish at spatial infinity).
Still it should always be possible to represent the
solution $A^a_i ({\bf r},t)$ by the ordered form
\begin{eqnarray}
\label{eq:sol}
A^a_i ({\bf r},t) =  W_i^{(0)a}({\bf r}) \hspace{-0.1cm} &+& \hspace{0.1cm}
\sum_{n=1}^{\infty}  \int d^2{\bf r}_1 \cdots d^2 {\bf r}_n
\psi^\dagger({\bf r}_1,t)\cdots\psi^\dagger({\bf r}_n,t)W^{(n)a}_i
({\bf r}; {\bf r}_1, T_{(1)},\cdots,{\bf r}_n, T_{(n)}) \nonumber \\
& & \cdot \psi({\bf r}_1,t) \cdots \psi ({\bf r}_n,t)
\end{eqnarray}
with suitable (matrix-valued) c-number functions $W^{(n)a}_i$
satisfying the symmetry restriction
\begin{equation}
\label{eq:con}
W^{(n)a}_i ({\bf r};\cdots,{\bf r}_p, T_{(p)},\cdots,{\bf r}_q, T_{(q)},\cdots
)=
W^{(n)a}_i ({\bf r};\cdots,{\bf r}_q, T_{(q)},\cdots,{\bf r}_p, T_{(p)},\cdots
)
\end{equation}
In Eq. (\ref{eq:sol}), the $T_{(p)}$ in $W^{(n)a}_i$ denote generator
matrices acting on $\psi({\bf r}_p,t)$ (and on $\psi^{\dagger} ({\bf r}_p,t)$
from the right)
and so satisfy $[T^a_{(p)},
T^b_{(q)}]=\delta_{pq} f^{abc} T^c_{(p)}$.
By inserting the form (5) into Eq. (3), we then obtain the conditions
that the functions $W_i^{(n)a}$ should satisfy. For the c-number contribution
$W_i^{(0)a}({\bf r})$, this consideration immediately tells us that it
is necessarily a pure gauge; we will set $W_i^{(0)a}({\bf r}) \equiv 0$
hereafter, as a part of our gauge choice. Now the rest of the functions
$W_i^{(n)a} ( {\bf r}; {\bf r}_1, T_{(1)}, \cdots ,{\bf r}_n, T_{(n) )}$,
or $W_i^{(n)a} ({\bf r}; 1,2, \cdots ,n)$ in short, are required to solve
the following equations iteratively:
\begin{eqnarray}
\label{eq:w}
\vspace{3cm}
\epsilon^{ij} \partial_i W^{(1)a}_j({\bf r}\hspace{-0.2cm}&;& \hspace{-0.2cm}1)
 + \frac{1}{2}\epsilon^{ij}f^{abc}
W^{(1)b}_i({\bf r};1)W^{(1)c}_j({\bf r};1)=
-\frac{i}{\kappa}T^a_{(1)}\delta({\bf r}-{\bf r}_1), \nonumber \\
\vspace{3cm}
\epsilon^{ij}  \partial_i W^{(2)a}_j({\bf r}\hspace{-0.2cm}&;&
\hspace{-0.2cm}1,2)
+\frac{1}{2}\epsilon^{ij}f^{abc} \left\{  W^{(1)b}_i({\bf r};1)W^{(1)c}_j({\bf
r};2)+ \left( W^{(1)b}_i({\bf r};1) +
 W^{(1)b}_i({\bf r};2) \right) W^{(2)c}_j({\bf r};1,2) \right.  \nonumber \\
& & \left. + W^{(2)b}_i  ({\bf r};1,2) \left( W^{(1)c}_j ({\bf r};1)
+W^{(1)c}_j({\bf r};2)
\right)
+2 W^{(2)b}_i({\bf r};1,2)W^{(2)c}_j({\bf r};1,2) \right\}=0, \nonumber  \\
\vspace{3cm}
\epsilon^{ij}\partial_iW^{(3)a}_j({\bf r}\hspace{-0.2cm}&;&
\hspace{-0.2cm}1,2,3) +\frac{1}{2} \epsilon^{ij} f^{abc}
\left\{ W^{(1)b}_i({\bf r};1)W^{(2)b}_j({\bf r};2,3)+ \cdots \right\}=0,
\nonumber \\
\cdots ,\rm{etc}.
\end{eqnarray}
Any particular set of solutions $W_i^{(n)a}$ satisfying these equations may
be chosen for our subsequent developments.

While the first of Eq. (\ref{eq:w}) has a simple solution (in the Coulomb
gauge),
\begin{equation}
\label{eq:lea}
W^{(1)a}_i({\bf r};1)=\frac{i}{\kappa} \epsilon^{ij} \partial_j
G({\bf r}-{\bf r}_1)T^a_{(1)},~~~~~
\left( G({\bf r}-{\bf r}_1)=\frac{1}{2\pi}\ln|{\bf r}-{\bf r}_1| \right)
\end{equation}
it is a nontrivial task to solve the rest of Eq. (\ref{eq:w}) and thereby
make available explicit forms for $W^{(n)a}_i$ with $n \geq 2$.
(Note that if one chose the axial gauge, all $W_i^{(n)a}$ with $n \neq 1$
could be taken to be zero; but, the expression for
$W_i^{(1)a} ({\bf r}; {\bf r}_1, T_{(1)} )$ assumes a very awkward form).
Despite
the lack of  explicit knowledge on $W_i^{(n)a}$ for $ n \geq 2$, the following
characterization of the
functions $W^{(n)a}_i$ will prove useful later. For arbitrary
$N (\geq 2)$, suppose we form the
c-number vector potential,
\begin{eqnarray}
\label{eq:pot}
{\cal A}^a_i({\bf r};1,\cdots,N-1 )\hspace{-0.3cm}&=&\hspace{-0.3cm}
\sum_{p=1}^{N-1}W^{(1)a}_i({\bf r};p)
+\sum_{p} \sum_{q  \neq p} W^{(2)a}_i ({\bf r};p,q) +
\sum_p \sum_{q \neq p} \sum_{r \neq p,q} W^{(3)a}_i ({\bf r};p,q,r) \nonumber
\\
&+&\cdots+\sum_p \sum_{q \neq p}\cdots\sum_{s \neq p,q,\cdots,s}W^{(N-1)a}_i
({\bf r};p,q,\cdots,s),
\end{eqnarray}
then, as can readily be verified using Eq. (\ref{eq:w}), this
provides the solution to the equations
\begin{equation}
\label{eq:agau}
\frac{1}{2} \epsilon^{ij} (\partial_i {\cal A}^a_j -\partial_j {\cal A}^a_i
+f^{abc} {\cal A}^b_i {\cal A}^c_j )=-\frac{i}{\kappa}
\sum_{p=1}^{N-1} T^a_{(p)} \delta^2({\bf r}-{\bf r}_p).
\end{equation}
So this potential defines a flat connection except at some discrete
points; ${\cal A}^a_i({\bf r};1,\cdots,N-1)$ describes the classical
non-Abelian Chern-Simons field configuration when $N-1$ non-Abelian
point sources are situated at positions ${\bf r} ={\bf r}_p
{}~~~(p=1,\cdots,N-1)$.
In fact one may completely specify the functions $W^{(n)a}_i$ through
Eqs. (\ref{eq:pot}), (\ref{eq:agau}) and (\ref{eq:con}). Also it will soon
become evident that if one is concerned with the $N$-particle sector of
the given system, one is in fact allowed to truncate the above series
expression (Eq. (5)) for the vector potentials $A_i^{a} ({\bf r},t)$
to $n=N-1$, all terms after $n=N-1$ in the sum not contributing at all.

The complicated expression obtained by inserting the form
(\ref{eq:sol}) into Eq. (\ref{eq:ham}) is our Hamiltonian.
The corresponding Heisenberg equations
of motion are
\begin{eqnarray}
i \frac{\partial \psi_\alpha ( {\bf r}, t) }{\partial t} & = &
  [  \psi_\alpha ( {\bf r}, t) , H  ] \nonumber \\
  &=& - \frac{1}{2m} ( \vec{D}^2 \psi )_\alpha ( {\bf r}, t) \nonumber \\
  & & + \frac{1}{2m} \int \! d^2 {\bf r}' \psi^\dagger ({\bf r}', t)
      \{ (\partial_i {\hspace{-0.1in}}^{^{^\leftarrow}} {'} - A^a_i ({\bf r}',
t) T'^a )
              K^b_i( {\bf r}', {\bf r}, t)_{\alpha \beta} T'^b
              \nonumber \\
  & & ~~~~~~~~~-  K^b_i ({\bf r}', {\bf r}, t)_{\alpha\beta} T'^b
      (\partial'_i + A^a_i ({\bf r}', t) T'^a )\}
      \psi ({\bf r}', t) \psi_\beta ( {\bf r}, t) \nonumber \\
  & & - \frac{1}{2m} \int \! d^2 {\bf r}' \psi^\dagger ({\bf r}', t)
      K^a_i ({\bf r}', {\bf r}, t)_{\alpha\beta} T'^a
      K^b_i ({\bf r}', {\bf r}, t)_{\beta\gamma} T'^b
      \psi ({\bf r}', t) \psi_\gamma ({\bf r}, t) \; , \nonumber \\
  & &  \label{eq:11}
\end{eqnarray}
where the generator matrices $T'^a$ act only on the fields with
coordinate $ {\bf r}'$, and we have defined the operators
$K^a_i( {\bf r}', {\bf r}, t)_{\alpha \beta}$ by the relation
\begin{eqnarray}
[ \psi_\alpha ({\bf r}_1, t), A^a_i ({\bf r}', t) ] &=&
       \{ \sum^\infty_{n=1} \int d^2 {\bf r}_2 \cdots d^2 {\bf r}_n
               \psi^\dagger ({\bf r}_2, t) n W^{(n)a}_i
               ({\bf r}'; 1,2,\cdots, n)_{\alpha\beta}
          \nonumber \\
  & & ~~~~~~~~~~\cdot \; \psi ({\bf r}_2, t) \cdots \psi ({\bf r}_n, t)
              \} \psi_\beta ({\bf r}_1, t) \nonumber \\
  & \equiv & K^a_i ({\bf r}', {\bf r}_1, t)_{\alpha\beta}
            \psi_\beta ({\bf r}_1, t) \; .
    \label{eq:12}
\end{eqnarray}
Here the indeces $\alpha,~\beta$ are used in association with
the matrices $T_{(1)}$. The last term in the right hand side
of Eq.~(\ref{eq:11}) corresponds to the quantum operator ordering
correction (an analogous term appears in the Abelian case as
well \cite{jp}); this ordering term is important to obain the correct
first-quantized description below.

Now let $| \Phi \rangle$ denote any Heisenberg-picture $N$-particle
state vector. Then the corresponding Schr\"{o}dinger wave function is
given by
\begin{equation}
\label{eq:13}
\Phi ({\bf r}_1, \alpha_1, \cdots, {\bf r}_N, \alpha_N ; t) =
 \langle 0 | \frac{1}{\sqrt{N !}} \psi_{\alpha_1} ({\bf r}_1, t) \cdots
         \psi_{\alpha_N} ({\bf r}_N, t) | \Phi \rangle \; ,
\end{equation}
where the nonrelativistic vacuum $| 0 \rangle$ satisfies the condition
$\psi_\alpha ({\bf r}, t) | 0 \rangle = 0$. Thanks to the operator
field equations (\ref{eq:11}), we may express the time derivative
of this $N$-body wave function as
\begin{equation}
\label{eq:14}
i \frac{\partial}{\partial t} \Phi ({\bf r}_1, \alpha_1, \cdots,
   {\bf r}_N, \alpha_N ; t) = A + B + C + D
\end{equation}
with
\begin{eqnarray}
A &=& \sum^N_{p=1} \langle 0 | \frac{1}{\sqrt{N!}}\psi_{\alpha_1}
   ({\bf r}_1, t) \cdots \psi_{\alpha_{p-1}} ( {\bf r}_{p-1}, t)
   \{ \, - \frac{1}{2m}
       (D_i D_i \psi )_{\alpha_p} ( {\bf r}_p, t)
   \, \} ~~~~~~~~~~~~~~~
   \nonumber \\
  & & \cdot \; \psi_{\alpha_{p+1}} ({\bf r}_{p+1}, t)
      \cdots \psi_{\alpha_N}  ({\bf r}_N, t) | \Phi \rangle \; ,
      \label{eq:15}
\end{eqnarray}
\begin{eqnarray}
B &=& \sum^N_{p=1} \langle 0 | \frac{1}{\sqrt{N!}}\psi_{\alpha_1}
   ({\bf r}_1, t) \cdots \psi_{\alpha_{p-1}} ( {\bf r}_{p-1}, t)
   \nonumber \\
  & & \cdot \; \{ \frac{1}{2m}
       \int \! d^2 {\bf r}' \psi^\dagger ({\bf r}', t)
          (\partial_i \hspace{-0.1in}^{^{^\leftarrow}} {'} - A^a_i ({\bf r}',
t) T'^a)
          K^b_i( {\bf r}', {\bf r}_p, t)_{\alpha_p \beta_p} T'^b
         \psi ({\bf r}', t) \psi_\beta({\bf r}_p, t) \}\nonumber \\
  & & \cdot \;
      \psi_{\alpha_{p+1}} ({\bf r}_{p+1}, t)
      \cdots \psi_{\alpha_N}  ({\bf r}_N, t) | \Phi \rangle \; ,
      \label{eq:16}
\end{eqnarray}
\begin{eqnarray}
C &=& \sum^N_{p=1} \langle 0 | \frac{1}{\sqrt{N!}}\psi_{\alpha_1}
   ({\bf r}_1, t) \cdots \psi_{\alpha_{p-1}} ( {\bf r}_{p-1}, t)
   \nonumber \\
  & & \cdot \; \{ \, -\frac{1}{2m}
       \int \! d^2 {\bf r}' \psi^\dagger ({\bf r}', t)
          K^b_i( {\bf r}', {\bf r}_p, t)_{\alpha_p \beta_p} T'^b
          (\partial_i' + A^a_i ({\bf r}', t) T^a)
       \psi ({\bf r}', t) \psi_{\beta_p} ({\bf r}_p, t)
      \, \}  \nonumber \\
  & & \cdot \; \psi_{\alpha_{p+1}} ({\bf r}_{p+1}, t)
      \cdots \psi_{\alpha_N}  ({\bf r}_N, t) | \Phi \rangle \; ,
      \label{eq:17}
\end{eqnarray}
\begin{eqnarray}
D &=& \sum^N_{p=1} \langle 0 | \frac{1}{\sqrt{N!}}\psi_{\alpha_1}
   ({\bf r}_1, t) \cdots \psi_{\alpha_{p-1}} ( {\bf r}_{p-1}, t)
   \nonumber \\
  & & \cdot \; \{ -\frac{1}{2m}
       \int \! d^2 {\bf r}' \psi^\dagger ({\bf r}', t)
          K^b_i( {\bf r}', {\bf r}_p, t)_{\alpha_p \beta_p} T'^b
          K^a_i ( {\bf r}', {\bf r}_p, t)_{\beta_p \gamma_p} T'^a
           \psi ({\bf r}', t) \psi_{\gamma_p}
            ({\bf r}_p, t) \, \}\nonumber \\
  & & \cdot \; \psi_{\alpha_{p+1}} ({\bf r}_{p+1}, t)
      \cdots \psi_{\alpha_N}  ({\bf r}_N, t) | \Phi \rangle \; ,
      \label{eq:18}
\end{eqnarray}
The matrix elements given here may be calculated by considering the
procedure of relocating the operators $\psi_{\alpha_1} ({\bf r}_1, t)
\cdots \psi_{\alpha_{p-1}} ({\bf r}_{p-1}, t)$, one by one, to the
right of those inside the curly brackets, while taking into account
various commutator terms generated in effecting the relocation. Then
all that has to be done is to collect those commutator terms ( here
remember that $\langle 0 | \psi^\dagger_\alpha ( {\bf r}, t) = 0$ )
and evaluate them using the relations like that in Eq. (\ref{eq:12}).
For more on this procedure, readers may consult Ref. \cite{lee}.
The result of this manipulation is
\begin{eqnarray}
A &=& -\frac{1}{2m} \sum^N_{p=1} \left\{ \partial^{(p)}_i +
     \left(\sum_{q_1 (<p)} W^{(1)a}_i ({\bf r}_p; q_1) +
            \sum_{q_1 (<p)} \sum_{q_2(<p) \neq q_1}
               W^{(2)a}_i ({\bf r}_p; q_1, q_2) + \cdots ~~~~~~~~~~~~~ \right.
\right.
               \nonumber \\
  & & \left. \left. \hspace{0.8cm} + \sum_{q_1 (<p)} \cdots \hspace{-0.4cm}
\sum_{q_{p-1} (<p) \neq q_1,
               \cdots, q_{p-2}} \hspace{-0.4cm} W^{(p-1)a}_i
               ({\bf r}_p; q_1, \cdots, q_{p-1} )
   \right)T^a_{(p)} \right\}^2 \Phi (1,\cdots,N;t), \hspace{0.4cm}
\label{eq:19}
\end{eqnarray}
\begin{eqnarray}
B &=&\hspace{-0.2cm} -\frac{1}{2m} \sum^N_{p=1} \sum_{q_1 (<p)}
\partial^{(q_1)}_i
       \left\{ \left( W^{(1)a}_i ({\bf r}_{q_1};p) +
               2 \hspace{-0.2cm} \sum_{q_2 (<p) \neq q_1} W^{(2)a}_i ({\bf
r}_{q_1};p,q_2)
               + \cdots  \right. \right. \nonumber \\
  & &  \left. \left. ~~~~ + (p-1) \sum_{q_2 (<p) \neq q_1}
            \cdots \hspace{0.8cm}\sum_{q_{p-1} (<p) \neq q_1, \cdots, q_{p-2}}
\hspace{-0.4cm}
               W^{(p-1)a}_i ( {\bf r}_p; p, q_2, \cdots, q_{p-1} )
           \right)
        T^a_{(q_1)} \right\} \Phi (1,\cdots,N;t)      \nonumber \\
  & & \hspace{-0.2cm} -\frac{1}{2m} \sum^N_{p=1} \sum_{q_1 (<p)} \left\{
    \left( \sum_{q_2 (<p)\neq q_1} W^{(1)a}_i ({\bf r}_{q_1}; q_2) +
            \sum_{q_2 (<p)\neq q_1} \sum_{q_3(<p) \neq q_1,q_2}
               W^{(2)a}_i ({\bf r}_{q_1}; q_2, q_3)  \right.  \right.
               \nonumber \\
  & &\left. ~~~~   + \cdots + \sum_{q_2 (<p)\neq q_1} \cdots
              \sum_{q_{p-1} (<p) \neq q_1, \cdots, q_{p-2}} W^{(p-2)a}_i
               ({\bf r}_{p}; q_2, \cdots, q_{p-1} )
     \right) T^a_{(q_1)} \nonumber \\
   & & ~~~~ \cdot \left( W^{(1)b}_i ({\bf r}_{q_1};p)+
              2 \sum_{q'_2 (<p) \neq q_1} W^{(2)b}_i ({\bf r}_{q_1};p,q'_2)
               + \cdots  + (p-1) \sum_{q'_2 (<p) \neq q_1}
             \right.  \nonumber \\
   & & ~~~~ \left. \left. \cdots
           \sum_{q'_{p-1} (<p) \neq q_1, q'_2,\cdots, q'_{p-2}}
           W^{(p-1)b}_i ( {\bf r}_{q_1}; p, q'_2, \cdots, q'_{p-1} )
   \right)T^b_{(q_1)} \right\} \Phi (1,\cdots,N;t), \nonumber \\
   & &   \label{eq:20}
\end{eqnarray}
\begin{eqnarray}
C &=& \hspace{-0.2cm} -\frac{1}{2m} \sum^N_{p=1} \sum_{q_1 (<p)}
       \left\{ \left( W^{(1)b}_i ({\bf r}_{q_1};p) +
               2 \sum_{q_2 (<p) \neq q_1} W^{(2)b}_i ({\bf r}_{q_1};p,q_2)
               + \cdots \right. \right. \nonumber \\
  & & \left. \left. ~~~~   + (p-1) \hspace{-0.2cm} \sum_{q_2 (<p) \neq q_1}
\cdots
             \hspace{-0.4cm}  \sum_{q_{p-1} (<p) \neq q_1, \cdots, q_{p-2}}
\hspace{-0.4cm}
               W^{(p-1)b}_i ( {\bf r}_{q_1}; p, q_2, \cdots, q_{p-1} ) \right)
          T^b_{(q_1)} \right\} \partial^{(q_1)}_i \Phi (1,\cdots,N;t)
\nonumber \\
  & & \hspace{-0.2cm}  -\frac{1}{2m} \sum^N_{p=1} \sum_{q_1 (<p)}
       \left\{ \left( W^{(1)b}_i ({\bf r}_{q_1};p) +
               2 \sum_{q_2 (<p) \neq q_1} W^{(2)b}_i ({\bf r}_{q_1};p,q_2)
               + \cdots  \right. \right. \nonumber \\
  & & \left. ~~~~ + (p-1) \sum_{q_2 (<p) \neq q_1} \cdots
               \sum_{q_{p-1} (<p) \neq q_1, \cdots, q_{p-2}}
               W^{(p-1)b}_i ( {\bf r}_{q_1}; p, q_2, \cdots, q_{p-1} )
            \right) T^b_{(q_1)}  \nonumber \\
  & & ~~~~ \cdot \left( \sum_{q'_2 (<p) \neq q_1}
                W^{(1)a}_i ({\bf r}_{q_1}; q'_2)
      + \sum_{q'_2 (<p) \neq q_1} \sum_{q'_3(<p) \neq q_1, q'_2}
               W^{(2)a}_i ({\bf r}_{q_1}; q'_2, q'_3) + \cdots
                \right. \nonumber \\
  & &  \left. \left. ~~~~ + \sum_{q'_2 (<p) \neq q_1} \cdots
          \sum_{q'_{p-1} (<p) \neq q_1, q'_2, \cdots, q'_{p-2}} W^{(p-2)a}_i
               ({\bf r}_{q_1}; q'_2, \cdots, q'_{p-1} )
     \right)T^a_{(q_1)} \right\} \Phi (1,\cdots,N;t),   \nonumber \\
       \label{eq:21}
\end{eqnarray}
\begin{eqnarray}
\hspace{-0.4cm} D &=& -\frac{1}{2m} \sum^N_{p=1} \sum_{q_1 (<p)}
       \left\{ \left( W^{(1)b}_i ({\bf r}_{q_1};p) +
               2 \sum_{q_2 (<p) \neq q_1} W^{(2)b}_i ({\bf r}_{q_1};p,q_2)
               + \cdots ~~~~~~~~~~~~~~~~~~~~~~~~~~~~~~~~~
                \right. \right. \nonumber \\
   & & \left. \left. ~~~~ + (p-1) \hspace{-0.4cm} \sum_{q_2 (<p) \neq q_1}
\hspace{-0.2cm} \cdots
             \hspace{-0.1cm}  \sum_{q_{p-1} (<p) \neq q_1, \cdots, q_{p-2}}
\hspace{-0.4cm}
               W^{(p-1)b}_i ( {\bf r}_{q_1}; p, q_2, \cdots, q_{p-1} )
              \right) T^b_{(q_1)} \right\}^2 \Phi (1,\cdots,N;t).
                    \label{eq:22}
\end{eqnarray}
We have here used the abbreviation $\Phi(1,\cdots,N;t)$ for the N-body wave
function.

Despite complicated-looking forms, the sum of the above four
contributions conspires to produce a remarkably simple expression:
\begin{eqnarray}
\lefteqn{A+B+C+D =} \nonumber \\
  & & -\frac{1}{2m} \sum^N_{p=1} \left\{ \partial^{(p)}_i +
     \left( \sum_{q_1 \neq p} W^{(1)a}_i ({\bf r}_p; q_1) +
            \sum_{q_1 \neq p} \sum_{q_2 \neq p, q_1}
               W^{(2)a}_i ({\bf r}_p; q_1, q_2) +  \cdots ~~~~~~~
            \right. \right.   \nonumber \\
  & & \left. \left.  + \sum_{q_1\neq p} \cdots
            \sum_{q_{p-1} \neq p, q_1, \cdots, q_{p-1}} W^{(N-1)}_i
               ({\bf r}_p; q_1, \cdots, q_{N-1} )
      \right) T^a_{(p)} \right\}^2 \Phi (1,\cdots,N;t) .  \label{eq:23}
\end{eqnarray}
To see that, it is useful to recast the expression in the right hand side
of Eq.~(\ref{eq:23}) using the relation
\begin{eqnarray}
\lefteqn{ \left( \sum_{q_1 \neq p} W^{(1)a}_i ({\bf r}_p; q_1 ) + \cdots +
       \sum_{q_1 \neq p} \cdots \sum_{q_{p-1} \neq p, q_1, \cdots, q_{p-1}}
       W^{(N-1)a}_i ({\bf r}_p; q_1, \cdots, q_{N-1} ) \right)}
       \nonumber \\
  &=&  \sum_{q_1 (< p)} \left( W^{(1)a}_i ({\bf r}_p; q_1) +
             \sum_{q_2(<p) \neq q_1} W^{(2)a}_i ({\bf r}_p; q_1, q_2) +
             \cdots   \right.         \nonumber \\
  & & \left. ~~~~~~~~~~ + \sum_{q_2 (<p) \neq q_1} \cdots
       \sum_{q_{p-1} (<p) \neq q_1, \cdots, q_{p-2}} W^{(p-1)a}_i
               ({\bf r}_p; q_1, \cdots, q_{p-1} )
     \right) \nonumber \\
  & & + \sum_{s (>p)} \left( W^{(1)a}_i ({\bf r}_p; s) +
               2 \sum_{q_2 (<s) \neq p} W^{(2)a}_i ({\bf r}_p;s,q_2)
               + \cdots  \right. \nonumber \\
  & & \left.  ~~~~~~~~~~~  + (s-1) \sum_{q_2 (<s) \neq p} \cdots
               \sum_{q_{s-1} (<s) \neq p,q_2, \cdots, q_{s-2}}
               W^{(s-1)a}_i ( {\bf r}_p;s, q_2, \cdots, q_{s-1} )
       \right)\; .\nonumber \\
  & &  \label{eq:24}
\end{eqnarray}
At the same time we write the summation $\sum^N_{p=1} \sum_{q_1 (< p)}$
appearing in $B$, $C$ and $D$ above as $\sum^N_{q_1=1}\sum_{p (>q_1)}$
and then rename the indices $(q_1, p)$ by
$(p,s)$; the result will be the expressions for $B,C$ and $D$, all of which
start with
the summation $\sum_{p=1}^{N} \sum_{s (> p) }^{}$. Then the remaining step is
just to make sure that those expressions indeed match various contributions
entering the right hand side of Eq. (23) when the relation (24) is used.
Based on this observation and our earlier identification (\ref{eq:pot}),
we thus have the $N$-body Schr\"{o}dinger equation in the
desired form
\begin{equation}
\label{sss}
i\frac{\partial}{\partial t} \Phi (1,\cdots,N;t ) =
-\frac{1}{2m}   \sum_{p=1}^{N} \left\{ \partial^{(p)}_i +
{\cal A}^a_i ({\bf r}_p;1, \cdots, p-1, p+1, \cdots,N)T^a_{(p)} \right\}^2
\Phi(1,\cdots,N;t).
\end{equation}
This is  the equivalent first-quantized description we have been after.

\section{DISCUSSIONS}
\hspace{1cm}
We have shown explicitly that the nonrelativistic quantum field theory
defined by the Hamiltonian (\ref{eq:ham}) with the operator constraints (3) is
equivalent to that defined by the many-body Schr\"{o}dinger equation
(25) with the c-number constraints (10).
Note that this complete equivalence between the two descriptions is
not entirely an expected result, considering the nonlinear nature of
our constraint equations.

The above $N$-body Schr\"{o}dinger equation in Eq. (25)
allows an immediate physical
interpretation: each particle is under the influence of a non-Abelian
vector potential produced by other particles (which act as non-Abelian
point sources in the sense of Eq. (\ref{eq:agau})). Especially, for
$N=2$, we know the {\it {exact}} expression for the vector potential in the
Coulomb gauge ( see Eq. (\ref{eq:lea})),
\begin{eqnarray}
{\cal A}^a_i ({\bf r}_1;2)& =& W^{(1)a}_i ({\bf r};2)   \nonumber \\
                          & =& \frac{i}{2\pi \kappa} \epsilon^{ij}
                           \frac{(x_1 -x_2)_j}{|{\bf r}_1 -{\bf r}_2 |}
\nonumber \\
                          & =&-{\cal A}^a_i ({\bf r}_2;1)
\end{eqnarray}
Using complex coordinates $z_p =x_p +i y_p $ and $\bar{z_p}=x_p -iy_p $
$(p=1, 2)$, the $2$-body Schr\"{o}dinger equation can then be conveniently
written as
\begin{equation}
\label{eq:two}
i\frac{\partial}{\partial t} \Phi(1,2;t) =
-\frac{1}{m} \sum_{p=1,2} \left( \bar{{\bf {\cal D}}}_{(p)} {\bf {\cal
D}}_{(p)}
+{\bf {\cal D}}_{(p)} \bar{{\bf {\cal D}}}_{(p)} \right)
\Phi(1,2;t),
\end{equation}
where
\begin{eqnarray}
\label{eq:der}
{\bf {\cal D}}_{(1)}& =&\frac{\partial}{\partial{z_1}} -\frac{1}{4\pi \kappa}
\frac{T^a_{(1)}T^a_{(2)}} {z_1 -z_2},~~~~~
\bar{{\bf {\cal D}}}_{(1)} =\frac{\partial}{\partial{\bar{z}_1}} +\frac{1}{4\pi
\kappa}
\frac{T^a_{(1)}T^a_{(2)}} {\bar{z}_1 -\bar{z}_2},  \nonumber \\
{\bf {\cal D}}_{(2)}& =&\frac{\partial}{\partial{z_2}} -\frac{1}{4\pi \kappa}
\frac{T^a_{(1)}T^a_{(2)}} {z_2 -z_1},~~~~~
\bar{{\bf {\cal D}}}_{(2)} =\frac{\partial}{\partial{\bar{z}_2}} +\frac{1}{4\pi
\kappa}
\frac{T^a_{(1)}T^a_{(2)}} {\bar{z}_2 -\bar{z}_1}.
\end{eqnarray}
Introducing a new $2$-body wave function $\Psi(1,2;t)$ by
\begin{equation}
\label{eq:new}
\Phi(1,2;t) =
\exp \left\{ -\frac{1}{4\pi\kappa} (\ln(z_1 -z_2)(\bar{z}_1 -\bar{z}_2))
T^a_{(1)}T^a_{(2)} \right\} \Psi(1,2;t),
\end{equation}
we can recast Eq. (\ref{eq:two}) as
\begin{equation}
\label{eq:hhh}
i\frac{\partial}{\partial t} \Psi(1,2;t) = -\frac{1}{m} \sum_{p=1,2}
\left( \bar{{\bf \bigtriangledown}}_{(p)} {\bf \bigtriangledown}_{(p)}
+{\bf \bigtriangledown}_{(p)} \bar{{\bf \bigtriangledown}}_{(p)} \right) \Psi
(1,2;t),
\end{equation}
where $ \bar{{\bf \bigtriangledown}}_{(1)}=\frac{\partial}{\partial
\bar{z_1}},~~
\bar{{\bf \bigtriangledown}}_{(2)} =\frac{\partial}{\partial \bar{z_2}},~~
{\bf \bigtriangledown}_{(1)}=
\frac{\partial}{\partial{z_1}} -\frac{1}{2\pi \kappa}
\frac{T^a_{(1)}T^a_{(2)}} {z_1 -z_2},$ and $
{\bf {\bigtriangledown}}_{(2)}=
\frac{\partial}{\partial{z_2}} -\frac{1}{2\pi \kappa}
\frac{T^a_{(1)}T^a_{(2)}} {z_2 -z_1}$.
This is precisely what one has in the holomorphic gauge \cite{vel,ltj}.
But, Eq. (\ref{eq:new}) being not a unitary transformation, the
normalization for $\Psi$ will be unconventional while $\Phi$ satisfies
the conventional normalization. We also note that the $2$-body scattering
based on Eq. (\ref{eq:two}) is not much different from the Abelian
Aharonov-B\"{o}hm scattering [6, 7, 9]. But, with three or more particles,
we do not even have the explicit expressions for ${\cal A}^a_i$ (in
the Coulomb gauge) and a full quantum treatment for this case
remains as a nontrivial future problem. Possibly one can look for
perturbative solutions to Eq. (7) (treating $\frac{1}{\kappa}$
as an expansion parameter) to find $W_i^{(n)a}$ for $n \geq 2$,
although an exact treatment should be much more desirable.

\vspace{1.5cm}
\hspace{-0.8cm}
Note: There appeared a very recent paper by Bak, Jackiw, and Pi
in which a systematic axial gauge treatment for the same model has been
given \cite{bak}.

\newpage
\section*{Acknowledgements}
We would like to thank R. Jackiw, S.-Y. Pi, D. Bak, T. Lee and P. Oh for
comments and useful discussions. This work was supported in part by the
Korea Science and Engineering Foundation through the Center for
Theoretical Physics, SNU and also by the Ministry of Education, Korea.

\end{document}